\documentclass{revtex4}

\newcommand{\beq}{\begin{eqnarray}}
\newcommand{\eeq}{\end{eqnarray}}
\newcommand{\bneq}{\begin{eqnarray*}}
\newcommand{\eneq}{\end{eqnarray*}}

\newcommand{\act}{\triangleright}
\newcommand{\cop}{\bigtriangleup}
\newcommand{\twist}{\mathcal{F}}
\newcommand{\scr}{\stackrel{*}{,}}
\newcommand{\cterm}[1]{1+\frac{{#1}_0}{\kappa}}
\newcommand{\pterm}[1]{\left (1+\frac{{#1}_0}{\kappa} \right )}

\begin{document}
\title{Differential Structure on $\kappa$-Minkowski Spacetime Realized
as Module of Twisted Weyl Algebra}

\author{Jong-Geon Bu}
\email{bjgeon@yonsei.ac.kr}
\author{Jae Hyung Yee}
\email{jhyee@yonsei.ac.kr}
\affiliation{Department of Physics, Yonsei University, Seoul 120-749,
Republic of Korea}
\author{Hyeong-Chan Kim}
\email{hckim@cjnu.ac.kr}
\affiliation{Division of Liberal Arts, Chungju National University, 
Chungju 380-702,
Republic of Korea}
\begin{abstract}
The differential structure on the $\kappa$-Minkowski spacetime
from Jordanian twist of Weyl algebra is constructed, 
and it is shown to be closed in 4-dimensions in contrast to the
conventional formulation. Based on this differential structure, 
we have formulated a scalar field theory in this $\kappa$-Minkowski
spacetime.
\end{abstract}
\maketitle

\section{Introduction}
$\kappa$-Minkowski spacetime consists of the noncommutative coordinates 
satisfying 
\beq
\label{eq:kappa}
[ x^0 \scr x^i] \equiv x^0 * x^i - x^i * x^0 = \frac{i}{\kappa} x^i.
\eeq
It was first introduced as translations in $\kappa$-Poincar\'e group, 
dual to the $\kappa$-Poincar\'e algebra~\cite{Lukierski, Majid}. 
It has attracted much interests as a realization of the so-called 
doubly special relativity (also known as deformed special 
relativity)~\cite{dsr}.
The differential structure~\cite{sitarz} 
and scalar field 
theory~\cite{Lukierskifield, rim, kappafieldsymmetry,kowalski,QuantumAspect}
have been formulated 
in this $\kappa$-Minkowski spacetime. Various physical implications 
have been investigated~\cite{kimvarious,various2}.
It is, however, known that the differential structure for 
this $\kappa$-Minkowski spacetime 
leads to the momentum 
space corresponding to 
a de Sitter section
in five dimensional flat space, 
and that there exist complex poles in the free scalar field propagator,
which implies the existence of unphysical ghost modes 
in this formulation~\cite{Lukierskifield}.

To circumvent the difficulties in understanding the physical implications 
of the five dimensional differential structure and the existence of 
complex modes in the free theories, 
the $\kappa$-Minkowski spacetime from 
twist deformation of underlying symmetry of the spacetime
were formulated~\cite{Majid_book}.
The virtue of the twist formulation is that the deformed symmetry algebra 
is the same as the original undeformed one and only the coproduct structure 
changes, leading to the same free field structure 
as the corresponding commutative field theory. 
Such twist formulation of noncommutative field theories 
was successfully applied to the case of the canonical 
noncommutative spacetime~\cite{Chaichian}.

The attempt to realize the $\kappa$-Minkowski spacetime by twisting Ponicar\'e
algebra is successful only for the light-cone $\kappa$-Minkowski 
spacetime~\cite{lightconeKappa}.
The realization of  the time-like $\kappa$-Minkowski spacetime by twist
was succeeded recently by enlarging the symmetry algebra of the spacetime 
to $igl(4,R)$ by the authors of \cite{IGL1} and  \cite{IGL2}.
The corresponding differential structure was constructed in~\cite{kim_diff},
which was shown to be closed in 4-dimensions contrary to 
the case of the conventional formulation which needs an extra fifth dimension. 
Based on this differential structure, scalar field theory is formulated 
in~\cite{kim_field}. 
Some physical properties of this twist realization
was also discussed in~\cite{IGL3}. 

The twist realization of the $\kappa$-Minkowski spacetime is also possible by 
twisting the Weyl and conformal algebras, which are smaller than $igl(4,R)$
and physically relevant,  by using the Jordanian 
twist~\cite{Herranz, Borowiec, Jordaniantwist, JordanianApp}.
One may also consider the chains of twists for classical Lie 
algebras~\cite{extendedjordanian}. It is the purpose of 
this paper to construct the differential structure of $\kappa$-Minkowski 
spacetime obtained by the Jordanian twist of the Weyl algebra, 
and to construct a free scalar field theory by using this twist 
approach. 

In sec.~\ref{twistWeyl}, we give a brief review of the Jordanian twist 
deformation of Weyl algebra, and find the correspoding $*$-product 
from the twist. 
We construct the differential structure of the $\kappa$-Minkowski spacetime, 
and find that it is closed in 4-dimensions in sec.~\ref{DiffCalc}. 
In sec.~\ref{action}, we construct a scalar field theory action, and 
discuss the physical implications of the theory in the last section. 

\section{Twisted Weyl (Hopf) algebra}
\label{twistWeyl}
In this section, we review the twist deformation 
of Weyl algebra by using the Jordanian 
twist~\cite{Majid_book, Herranz, Borowiec}. Given a Hopf algebra, 
one can define another (deformed) Hopf algebra using twist 
element $\mathcal{F}$ which obeys 
\beq
\label{eq:twistcondition}
(\twist \otimes 1)(\cop \otimes id) \twist &=& 
(1 \otimes \twist )(id \otimes \cop) \twist,\\
(\epsilon \otimes 1) \twist &=& (1 \otimes \epsilon)\twist.
\eeq
The product and the counit of the deformed Hopf algebra do not change, 
but the coproduct is deformed according to 
\beq
\label{deformedCop}
\cop_t Y = \twist \cop_0 Y \twist^{-1},
\eeq
where $Y$ represents generators of the original algebra 
and $\cop_0 Y = Y \otimes 1 + 1 \otimes Y$.
Along with this deformation, module algebra must also be deformed 
for the algebra to act covariantly, that is, 
the action of the generators of twist deformed Hopf algebra 
must be of the same form as that of the original algebra  
\beq
Y \act m(f \otimes g) = m ( \cop Y \act f \otimes g) 
= m (Y^{(1)} \act f \otimes Y^{(2)} \act g ),
\eeq 
where $m$ denotes the product of the module algebra  
and we use the Swedler notation $\cop Y = Y^{(1)} \otimes Y^{(2)}$. 
The deformed module algebra is defined by replacing the product $m$ 
by the new $*$-product, $m_*$, 
\beq
\label{eq:starpro}
f * g = m_{*} ( f \otimes g) = \mu ( \twist^{-1} \act f \otimes g).
\eeq 

We can apply this deformation to  Weyl (Hopf) algebra, 
which is generated by 11 generators, $P_0$, $P_i$, $M_i$, $N_i$, and  $D$,
each correspoding to those of time translation, space translation, 
rotation, boost and dilatation, respectively. 
These generators satisfy~\cite{CFT} 
\begin{eqnarray}
\begin{array}{ll}
[D, P_\mu]= i P_\mu, &
[D, M_i] = 0 = [D, N_i], \\

[P_0, M_i ] = 0, &
[P_0, N_i ] = i P_i,\\

[P_i, M_j ] = i \epsilon_{ijk} P_k, &
[P_i, N_j ] = -i \eta_{ij} P_0,\\

[M_i, M_j]= i \epsilon_{ijk} M_k, &
[M_i, N_j ] = i \epsilon_{ijk} N_k,\\

[N_i, N_j] = -i \epsilon_{ijk} M_k.
\end{array}
\end{eqnarray}
The greek indices $\mu$ and $\nu$ run over  0, 1, 2, 3, 
Latin indices $i$, $j$ and $k$ run over  1, 2, 3, 
and $\eta_{\mu \nu} = {\rm diag}(+1, -1, -1, -1)$. The repeated index 
implies the summation over that index. 

Weyl algebra has a representation on the space of 
coordinates $x^\mu$:
\beq
\label{eq:AonC}
\begin{array}{ll}
D \act x^\rho = - i x^\rho, &P_\mu \act x^\rho = - i \eta^{\rho}_\mu, \\
M_i \act x^0 = 0, & M_i \act x^l  = - i \epsilon^{l}_{\phantom{l}im} x^m,\\
N_i \act x^0 = -i \eta_{i \rho}x^\rho, &
N_i \act x^l = i \eta^{l}_{\phantom{l}i} x^0.
\end{array}
\eeq

As a twist element, we choose  
\beq
\label{eq:twist1}
\twist 
= \exp \left ( -i D \otimes \ln \left (1 + \frac{P_0}{\kappa}\right )\right ),
\eeq
which satisfies Eq.~(\ref{eq:twistcondition})  
and known as a Jordanian twist~\cite{Jordaniantwist,Herranz,Borowiec}.
Coproduct of the twisted Weyl algebra is obtained from Eq.~(\ref{deformedCop}): 
\beq
\cop_t D &=& 1 \otimes D + D \otimes \frac{1}{1 + \frac{P_0}{\kappa}},\\
\label{eq:deformedCopforp}
\cop_t P_\mu &=& P_\mu \otimes \left( 1 + \frac{P_0}{\kappa} \right ) 
+  1 \otimes P_\mu, \\
\cop_t M_i &=& M_i \otimes 1  + 1 \otimes M_i, \\
\cop_t N_i &=& N_i \otimes 1 + 1 \otimes N_i 
+ D \otimes \frac{\frac{P_i}{\kappa}}{1+\frac{P_0}{\kappa}}.
\eeq
In twisted Weyl algebra, the spatial rotations are undeformed, thereby retaining
the rotational symmetry of the 3-dimensional space.

As a module algebra, the space of the functions 
also has to be deformed. The product of this algebra is replaced 
by the $*$-product defined by Eq.~(\ref{eq:starpro}). Using this 
$*$-product, the product of two coordinates becomes
\beq
\label{eq:prd2x}
x^\mu * x^\nu = x^\mu x^\nu - \frac{i}{\kappa} \eta^{0 \nu} x^\mu,
\eeq
and the commutation relations between coordinates become 
\beq
\label{eq:kappa-rel}
[x^0 \scr x^i ] &\equiv& x^0 * x^i - x^i * x^0  = \frac{i}{\kappa} x^i, 
\,\, [x^i \scr x^j ] = 0,
\eeq
which is the commutation relation for $\kappa$-Minkowski spacetime,  
Eq.~(\ref{eq:kappa}). 

For later convenience, we calculate the $*$-product of 
two exponential functions,
\beq
e^{ip\cdot x} * e^{iq\cdot x} = e^{i (p + q + \frac{q_0}{\kappa} p)\cdot x}
\eeq
where $p \cdot x = p_\mu x^\mu$. 
As expected, this implies the addition of momenta described 
by the coproduct of $P_\mu$, Eq.~(\ref{eq:deformedCopforp}).
This $*$-product corresponds to the time-left ordering, 
\beq
:e^{ip\cdot x}: = e^{ip_0 x^0} * e^{ip_i x^i},
\eeq
of the exponential kernel function in the conventional approach of 
the $\kappa$-Minkowski spacetime. 

In the case of  twist deformation 
of $igl(4,R)$~\cite{IGL1, IGL2}, there exists 
a free parameter in the twist element. Depending on the value of the parameter, 
the resultant $*$-product corresponds to a specific ordering prescription 
of the exponential function in the conventional formulation. 
There also exists such a freedom in the case of the Jordanian 
twist. If one chooses the twist element,
\beq
\twist_2 = \exp \left(
-i \ln\left(1 -\frac{P_0}{\kappa}\right ) 
\otimes D \right ),
\eeq
instead of~(\ref{eq:twist1}), the resultant $\star$-product
becomes 
\beq
e^{i p \cdot x } \star e^{i q \cdot x}
 = e^{i ( p + q - \frac{p_0}{\kappa} q)\cdot x},
\eeq
which corresponds to the time-right ordering, 
\beq 
:e^{i p \cdot x}: = e^{i p_i x^i } \star e^{i p_0 x^0},
\eeq
in the conventional formulation. 

It is noted that with the twist~(\ref{eq:twist1}) we have 
\beq
x^0 * x^0 = (x^0)^2 - \frac{i}{\kappa} x^0 \neq (x^0)^2, \nonumber
\eeq
and  
\beq
e^{i p_0 x^0} * e^{i q_0 x^0} = e^{i (p_0 + q_0 + \frac{p_0 q_0}{\kappa}) x^0}
\neq e^{i (p_0 + q_0) x^0}, \nonumber
\eeq
in contrast to the conventional formulation and the $igl(4,R)$ 
twist formulation. 

\section{Differential Structure}
\label{DiffCalc}
In Ref.~\cite{sitarz}, a bicovariant differential calculus~\cite{Woronowicz}
was constructed on the $\kappa$-Minkowski spacetime based on the 
$\kappa$-Poincar\'e algebra. 
In the case of this conventional $\kappa$-Minkowski
spacetime based on the $\kappa$-Poincar\'e algebra, 
it turns out that there does not exist 4-D differential calculus 
which is Lorentz covariant, but one needs to introduce an extra fifth dimension.

In a similar way, we obtain the differential calculus 
on $\kappa$-Minkowski spacetime 
of the last section. 
The external derivative 
$d$ is demanded to satisfy the Leibniz rule:
\beq
\label{eq:leibniz}
d ( f*g) = df * g + f * dg.
\eeq
For the generators $Y$ of the Weyl algebra, the action on the differential 
algebra may be postulated as 
\beq
Y \act d x^\mu &=& d ( Y \act x^\mu ), \nonumber \\
Y \act x^\mu * dx^\nu 
&=& (Y^{(1)} \act x^\mu ) * (Y^{(2)} \act dx^\nu). \nonumber
\eeq

For the twisted Weyl algebra, we obtain the following identities 
from the representation~(\ref{eq:AonC}):
\bneq
D \act [x^0 \scr dx^\mu] 
&=& -2 i [x^0 \scr dx^\mu] - \frac{1}{\kappa} dx^\mu, \\
D \act [x^l \scr dx^\mu] &=& -2 i [x^l \scr dx^\mu], \\
M_i \act [x^0 \scr dx^0 ] &=& 0,\\
M_i \act [x^0 \scr dx^l ] &=& i \epsilon^{l}_{\phantom{l}{im}} 
[x^0 \scr dx^m], \\
M_i \act [x^l \scr dx^0 ] &=& i \epsilon^{l}_{\phantom{l}{im}} 
[x^l \scr dx^m],\\
M_i \act [x^l \scr dx^m ] &=& i \epsilon^{m}_{\phantom{m}{in}} [x^l \scr dx^n] 
+ i \epsilon^l_{\phantom{l}{in}}[x^n \scr dx^m],\\
N_i \act [ x^0 \scr dx^0 ] &=& -i \eta_{il} [x^l \scr dx^0] 
- i \eta_{il} [x^0 \scr dx^l],\\
N_i \act [ x^0 \scr dx^l]&=& - i \eta_{im} [x^m \scr dx^l] 
+ i \eta^l_{\phantom{l}i} [x^0 \scr dx^0],\\
N_i \act [x^l \scr dx^0] &=& i \eta^{l}_{\phantom{l}i} [x^0 \scr dx^0]
-i \eta_{im} [ x^l \scr dx^m]\\ 
& & + \frac{1}{\kappa} \eta^{l}_{\phantom{l}i} dx^0, \\
N_i \act [x^l \scr dx^m ] &=& i \eta^{l}_{\phantom{l}i}[x^0 \scr dx^m]
+ i \eta^m_{\phantom{m}i}[x^l \scr dx^0]\\ 
& &+ \frac{1}{\kappa} \eta^{l}_{\phantom{l}i} dx^m.
\eneq
We find that  these identities are satisfied if we demand 
\beq
\label{eq:diff_calculus}
\begin{array}{rl}
[x^0 \scr dx^\mu] &= \frac{i}{\kappa} dx^\mu, \\ 
\,[x^i \scr dx^\mu] &= 0,
\end{array}
\eeq
which is similar to the differential structure of 
$igl(4,R)$ twist formulation, except that $[x^0 \scr dx^0]$
does not vanish. 
If one tries similar differential structure in the $\kappa$-Minkowski
spacetime based on the $\kappa$-Poincar\'e algebra as (\ref{eq:diff_calculus}),
they do not satisfy the mixed Jacobi identity~\cite{sitarz}.  
However, it is easy to show that the commutation 
relations (\ref{eq:diff_calculus}) is consistent with 
the mixed Jacobi identity: 
\beq
[x^\mu \scr [x^\nu \scr dx^\rho]] + [ x^\nu \scr [dx^\rho \scr x^\mu ] ]
 + [dx^\rho \scr [x^\mu \scr x^\nu]] = 0, 
\eeq
and the commutation relations for the $\kappa$-Minkowski spacetime, 
Eq.~(\ref{eq:kappa-rel}).

Eq.~(\ref{eq:diff_calculus}) defines the differential structure
of the $\kappa$-Minkowski spacetime from the Jordanian twist 
of Weyl algebra. 
Contrary to the conventional $\kappa$-Minkowski spacetime 
in which only covariant 5-$D$ differential calculus exists,
the differential calculus is closed in 4-dimensions and 
is covariant under the twisted Weyl algebra.

We may define partial derivative $\partial_\mu$ 
as
\beq 
\label{eq:relwithpartial}
df = \partial_\mu f * dx^\mu.
\eeq
To find the properties of the partial derivative, it is sufficient 
to examine the derivatives of the exponential function. 
From Eq.~(\ref{eq:leibniz}) and Eq.~(\ref{eq:diff_calculus})
we find
\beq
\label{eq:cmtr_exp}
dx^\mu * e^{iq \cdot x} 
= \left ( 1 + \frac{q_0}{\kappa} \right ) e^{iq \cdot x} * dx^\mu,
\eeq
and 
\beq 
\label{eq:diff_exp}
de^{ip \cdot x} = ip_\mu e^{ip \cdot x} * dx^\mu. 
\eeq

Comparing  Eq.~(\ref{eq:relwithpartial}) and Eq.~(\ref{eq:diff_exp}),
we find that  $\partial_\mu$ acts like an ordinary partial 
derivative,
\beq
\partial_\mu e^{ip \cdot x} = ip_\mu e^{ip \cdot x}.
\eeq 
However, this partial derivative $\partial_\mu$ does not obey the ordinary 
Leibniz rule, but satisfies   
\beq
\label{eq:Leibniz}
\partial_\mu ( f * g) = \partial_\mu f * g + f * \partial_\mu g 
- \frac{i}{\kappa} \partial_\mu f * \partial_0 g.
\eeq

\section{Action for scalar field theory}
\label{action}
In this section we first consider  the Fourier transfomation of a scalar field,
define an adjoint derivative, and then 
write down the action for the massless scalar field theory invariant under 
the twisted Weyl algebra. 

We define the Fourier transformation of a scalar field as
\beq
\label{eq:Fourier}
\phi(x) &=& \int d\mu(p) e^{ip \cdot x} \tilde{\phi}(p) ,
\eeq
where $d\mu(p)$ is the integration measure to be determined. 
The inverse Fourier transformation 
has to be  defined using $*$-product and the conjugate of 
exponential function. We demand that the conjugate of 
the exponential function obeys the relations,
\beq
(e^{ip \cdot x} * e^{iq \cdot x})^\dagger 
&=& (e^{iq \cdot x})^\dagger * (e^{ip \cdot x})^\dagger, \nonumber\\
e^{ip \cdot x} * (e^{ip \cdot x})^\dagger &=& 1 
= (e^{ip \cdot x})^\dagger * e^{ip \cdot x},\nonumber\\
((e^{ip \cdot x})^\dagger)^\dagger &=& e^{ip \cdot x}, \nonumber
\eeq
and find that these relations are satisfied if we define the
conjuagation of exponetial function as
\beq
(e^{ip \cdot x})^\dagger 
= \exp \left ( i \left ( -\frac{p_\mu}{1
+ \frac{p_0}{\kappa}} \right ) x^\mu \right ) .
\eeq

From this, we define the deformed antipode $S_t$ of $P_\mu$ as 
in Ref.~\cite{kowalski}: 
\beq
S_t(P_\mu) = \frac{- P_\mu}{1 + \frac{P_0}{\kappa}}.
\eeq 
We find that this definition obeys the relation,
\beq
\cdot (S \otimes id) \cop = \cdot (id \otimes S) \cop = \eta \epsilon,
\eeq
for  the antipode~\cite{Majid_book} and is the same as that in~\cite{Borowiec}.

Using this conjugate of exponential function, we define 
the delta function: 
\beq
\int_x (e^{ip \cdot x})^\dagger * e^{iq \cdot x} = (2 \pi)^4 \pterm{p}
\delta^{(4)} (q-p),
\eeq 
where $\int_x \equiv \int d^4x$.
We now can define the inverse Fourier transformation as
\beq
\label{eq:inverseF}
\tilde{\phi}(p) &=& \int_x (e^{ip \cdot x})^\dagger * \phi(x). 
\eeq
From the definition of Fourier transformation and the delta function, 
we find the invariant measure
\beq
d \mu(p)= \frac{d^4 p}{(2 \pi)^4 \left ( 1+ \frac{p_0}{\kappa}\right )}.
\eeq

Adjoint derivative $\partial^\dagger$ is defined as   
\beq
\int_x \phi(x)* \partial_\mu \phi(x) 
= \int \partial^\dagger_\mu \phi(x) * \phi(x).
\eeq
It can be shown that 
\beq
\partial^\dagger_\mu e^{ip \cdot x} 
= i \left ( - \frac{p^\mu}{1 + \frac{p_0}{\kappa}} \right )e^{ip \cdot x}, 
\quad (\partial_\mu e^{ip \cdot x})^\dagger 
= - \partial^\dagger_\mu (e^{ip \cdot x})^\dagger.
\eeq

The action for massless scalar field can be written in analogy with that of 
the commutative spacetime as 
\beq
S = \int_x (\partial_\mu \phi(x))^\dagger  * \partial^\mu \phi(x).
\eeq
Using the Fourier transformation defined above, we write 
the action in  momentum space
\beq
S = \int_p \tilde{\phi}^\dagger(p)~ p_\mu p^\mu~ \tilde{\phi}(p),
\eeq
which is consistent with the dispersion relation discussed in~\cite{Borowiec}.
Thus it implies that in momentum space the free field structure 
of the $\kappa$-Minkowski spacetime is the same as 
that of the corresponding commutative theory 
except for the measure factor as expected in the twist formulation. 

\section{Discussion}
We have constructed a differential structure on $\kappa$-Minkowski 
spacetime from twisting the Weyl symmetry group~\cite{Borowiec},
and have shown that the differential structure is closed in 4-dimensions
as in the case of the twist formulation based on  $IGL(4,R)$ 
group~\cite{kim_diff}.
Based on this differential structure,
we have formulated a free scalar field theory in this noncommutative
spacetime. As in the case of the $IGL(4,R)$ twist, 
the dispersion relation is the same as the correponding commutative theory, 
and the theory has the same free field structure as in  the commutative case. 
Thus one may avoid various difficulties encountered in the conventional 
formulation of the $\kappa$-Minkowski spacetime based on the 
$\kappa$-Poincar\'e algebra, and study the effects of the 
$\kappa$-deformation more easily using this twist formulation. 

It is noted that, contrary to other approaches, we have 
$x^0 * x^0 = (x^0)^2 -\frac{i}{\kappa} x^0$ and 
$[x^0 \scr dx^0] = \frac{i}{\kappa} dx^0$. One may relate
the action of translation generator $P_\mu$ on the space of functions 
to the external derivative $d$ as
\beq 
df = (iP_\mu \act f) * dx^\mu.
\eeq
With this relation, Eq.~(\ref{eq:Leibniz}) is consistent 
with the action on the deformed space of functions of coordinates 
for translation generator $P_\mu$ 
of the twisted Weyl algebra, whose coproduct is given by
Eq.~(\ref{eq:deformedCopforp}),
\beq 
\partial_\mu (f * g) = i P_\mu \act m_*( f \otimes g).
\eeq
It is noted that the external derivative $d$ satisfies the Leibniz rule 
even though partial derivative $\partial_\mu$ does not. This comes 
from the commutation relation between $dx^\mu$ and the exponential 
function, Eq.~(\ref{eq:cmtr_exp}):
\beq
d(f*g) &\equiv& \partial_\mu ( f* g) * dx^\mu \nonumber \\
&=&\left [ \partial_\mu f * \left ( g
 - \frac{i}{\kappa} \partial_0g \right ) \right ]* dx^\mu 
+ ( f * \partial_\mu g ) * dx^\mu \nonumber \\
&=& (\partial_\mu f * dx^\mu) * g + f * (\partial_\mu g * dx^\mu) \nonumber \\
&=& df * g + f * dg. \nonumber
\eeq

As in other twist formulations, the intrinsic effects of 
such $\kappa$-deformation would appear as a result of interactions. 
As an interaction in the Weyl symmetric theories 
one would naturally consider the $\phi^4$-interaction, 
\beq
\frac{\lambda}{4!} \int_x 
\phi(x) * \phi(x)* \phi(x) * \phi(x). 
\eeq
This interaction term becomes, in momentum space, 
\beq
\label{eq:int}
\frac{\lambda}{4!} \, (2\pi)^4 \int_{klpq} 
\tilde{\phi}(k)\tilde{\phi}(l)\tilde{\phi}(p) \tilde{\phi}(q)
\delta (k_{lpq} + l_{pq} + p_q + q ),
\eeq
where $k_{lpq}=k(\cterm{l})(\cterm{p})(\cterm{q})$, 
$l_{pq}= l (\cterm{p})(\cterm{q})$, $p_q = p (\cterm{q})$. 

By change of variables,
$p' =  S_t(p) = -\frac{p}{\cterm{p}}$, $q'=S_t(q)= -\frac{q}{\cterm{q}}$,
using the deformed antipode $S_t$ of $P_\mu$,
we can rewrite Eq.~(\ref{eq:int}) as
\beq
\begin{array}{l}
\frac{\lambda}{4!} \, (2\pi)^4 \int_{klpq} 
\tilde{\phi}(k)\tilde{\phi}(l)\bar{\phi}(p) \bar{\phi}(q)\\
\quad \times \delta\left ( (k\pterm{l}+l)-(q\pterm{p}
+p) \right ),
\end{array}
\eeq
where $\bar{\phi}(q) = \frac{\tilde{\phi}(S_t(q))}{\pterm{q}^{-1}}$.
If we interprete this interaction term as scattering process
of two incoming particle $\bar{\phi}$ into two outgoing $\tilde{\phi}$, 
then the delta-function in the integrand guarantees  that
the total  momentum of in-particles is equal to that of out-particles. 

It would be interesting to investigate the physical effect of this interaction
and study the possibility of describing the
Planck scale physics by such $\kappa$-deformation of spacetime.

For Weyl symmetry, we have to consider only the massless 
theory. It would be interesting to investigate 
the possibility of Weyl symmetry breaking through quantum corrections.

\begin{acknowledgments}
This work was supported in part by the Korea Science and Engineering
Foundation(KOSEF) grant through the Center for Quantum spacetime(CQUeST)
of Sogang University with grant number R11-2005-021.
\end{acknowledgments}

\end{document}